\documentclass[prl,letter,twocolumn,superscriptaddress]{revtex4-2}
\usepackage[utf8]{inputenc}
\usepackage[american,]{babel}
\usepackage[T1]{fontenc}
\usepackage[pdftex]{graphicx}  
\usepackage{xcolor}
\usepackage{dcolumn}
\usepackage{bm}
\usepackage{amsmath,amsthm,amssymb}
\usepackage{hyperref}
\hypersetup{colorlinks,bookmarksopen,bookmarksnumbered,
    citecolor=blue,
    linkcolor=blue,
    pdfstartview=false,
    urlcolor=blue}
\usepackage{braket}
\usepackage{wrapfig}
\usepackage{soul}
\usepackage{mathtools}
\usepackage{float}

\renewcommand{\vec}{\mathbf}

\begin{document}
\title{Entanglement Growth and Minimal Membranes in $(d+1)$ Random Unitary Circuits}

\author{Piotr Sierant}
\affiliation{ICFO-Institut de Ci\`encies Fot\`oniques, The Barcelona Institute of Science and Technology, Av. Carl Friedrich Gauss 3, 08860 Castelldefels (Barcelona), Spain}
\author{Marco Schir\`o}
\affiliation{JEIP, USR 3573 CNRS, Coll\`{e}ge de France, PSL Research University, 11 Place Marcelin Berthelot, 75321 Paris Cedex 05, France}
\author{Maciej Lewenstein}
\affiliation{ICFO-Institut de Ci\`encies Fot\`oniques, The Barcelona Institute of Science and Technology, Av. Carl Friedrich Gauss 3, 08860 Castelldefels (Barcelona), Spain}
\author{Xhek Turkeshi}
\affiliation{JEIP, USR 3573 CNRS, Coll\`{e}ge de France, PSL Research University, 11 Place Marcelin Berthelot, 75321 Paris Cedex 05, France}
\date{\today}
\begin{abstract}
Understanding the nature of entanglement growth in many-body systems is one of the fundamental questions in quantum physics. 
Here, we study this problem by characterizing the entanglement fluctuations and distribution of $(d+1)$ qubit lattice evolved under a random unitary circuit.
Focusing on Clifford gates, we 
perform extensive numerical simulations of random circuits in $1\le d\le 4$ dimensions. 
Our findings demonstrate that properties of growth of bipartite entanglement entropy are characterized by the roughening exponents of a $d$-dimensional membrane in a $(d+1)$ elastic medium.
\end{abstract}

\maketitle
Characterizing entanglement evolution in many-body quantum systems is an essential problem in statistical mechanics, condensed matter theory, and high-energy physics~\cite{laflorencie2016quantum,amico2008entanglement,potter2022entanglementdynamicsin}. 
For instance, it has profound implications for quantum thermalization~\cite{deutsch1991quantumstatisticalmechanics,srednicki1994chaosandquantum,pappalardi2022eigenstatethermalizationhypothesis,gogolin2016equilibrationthermalizationand,dalessio2016fromquantumchaos,zhang2015thermalization,ho2017entanglement,znidari2020entanglementgrowthin} and its breakdown in disordered systems~\cite{chiara2006entanglement,znidaric2008manybody,bardarson2012unboundedgrowthof,huse2014phenomenology,abanin2019colloquium,serbyn2013universalslowgrowth,nandkishore2015manybodylocalizationand,Alet18, sierant22challenges}. At the same time, it provides new perspectives on quantum chaos arising in semiclassical and long-range systems~\cite{schachenmayer2013entanglement,hauke2013spread,lerose2020origin,lerose2020bridging,pappalardi2018scrambling} or holographic models~\cite{abajoarrastia2010holographic,hubeny2007a,liu2014entanglementgrowthduring,liu2014entanglementtsunamiuniversal,roberts2015localizedshocks,swingle2012entanglementrenormalizationand,hayden2007blackholesas,sekino2008fastscramblers}.
Yet, understanding entanglement dynamics is of outstanding difficulty due to the nonlocal nature of the quantum correlations. Insights in one-dimensional systems come from conformal field theory and integrability~\cite{calabrese2004entanglement,calabrese2005evolutionofentanglement,calabrese2009entanglemententropyand,calabrese2016quantum,asplund2015entanglement,fagotti2008evolution,peschel2009reduced,alba2017entanglement,alba2018entanglement,alba2019quantum,alba2019quantum1}, exactly solvable dual-unitary dynamics~\cite{bertini2019entanglementspreadingin,bertini2020operatorentanglementin,gopalakrishnan2019unitarycircuitsof,piroli2020exactdynamicsin,reid2021entanglementbarriersin,jonay2021triunitaryquantumcircuits,bertini2022entanglementnegativityand}, and extensive numerical investigations~\cite{kim2013ballisticspreadingof,lauchli2008spreadingofcorrelations,rigol2008thermalizationandits,oliveira2007genericentanglementcan,znidaric2008exactconvergencetimes,dahlsten2014entanglementtypicality,schollwock2011the,verstraete2008matrixproductstates}.
These results motivate the search for a phenomenological understanding of quantum information scrambling in generic many-body systems.

Random unitary circuits are ideal candidates for this purpose~\cite{fisher2023randomquantumcircuits}. In suitable limits, Ref.~\cite{nahum2017quantumentanglementgrowth,zhou2019emergentstatisticalmechanics,zhou2020entanglementmembranein} demonstrated that the evolution of entanglement entropy of $(1+1)D$ circuits maps to the properties of a line defect in a two-dimensional elastic manifold. 
We note that those insights go beyond the leading order contribution to the entanglement entropy growth and are reflected in \emph{structural} aspects of entanglement propagation. A non-trivial verification of this fact is the Kardar-Parisi-Zhang (KPZ) universal growth of entanglement fluctuations in $(1+1)D$ Clifford and Haar random unitary circuits~\cite{nahum2017quantumentanglementgrowth,zhou2019emergentstatisticalmechanics,zhou2020entanglementmembranein}. 
Inspired by these results, the authors of Ref.~\cite{nahum2017quantumentanglementgrowth} conjectured that generic $(d+1)$ dimensional systems correspond to a $d$-dimensional \emph{membrane} in a $(d+1)$ elastic medium. This defect is fixed by the \emph{minimal cut} of entanglement bonds across space-time, resulting, for random unitary circuits, in the entanglement growth being governed by exponents of a membrane pinned by the disorder~\cite{nattermann1985isingdomainwall,huse1985pinningandroughening,ferrero2021creepmotionof,ferrero2021creepmotionof,wiese2022theoryandexperiments}.
Without any numerical and analytical results for $d>1$, it remains unclear whether the minimal membrane picture indeed provides insights into the structure of entanglement evolution in higher dimensional systems.

\begin{figure}[t!]
    \centering
    \includegraphics[width=\columnwidth]{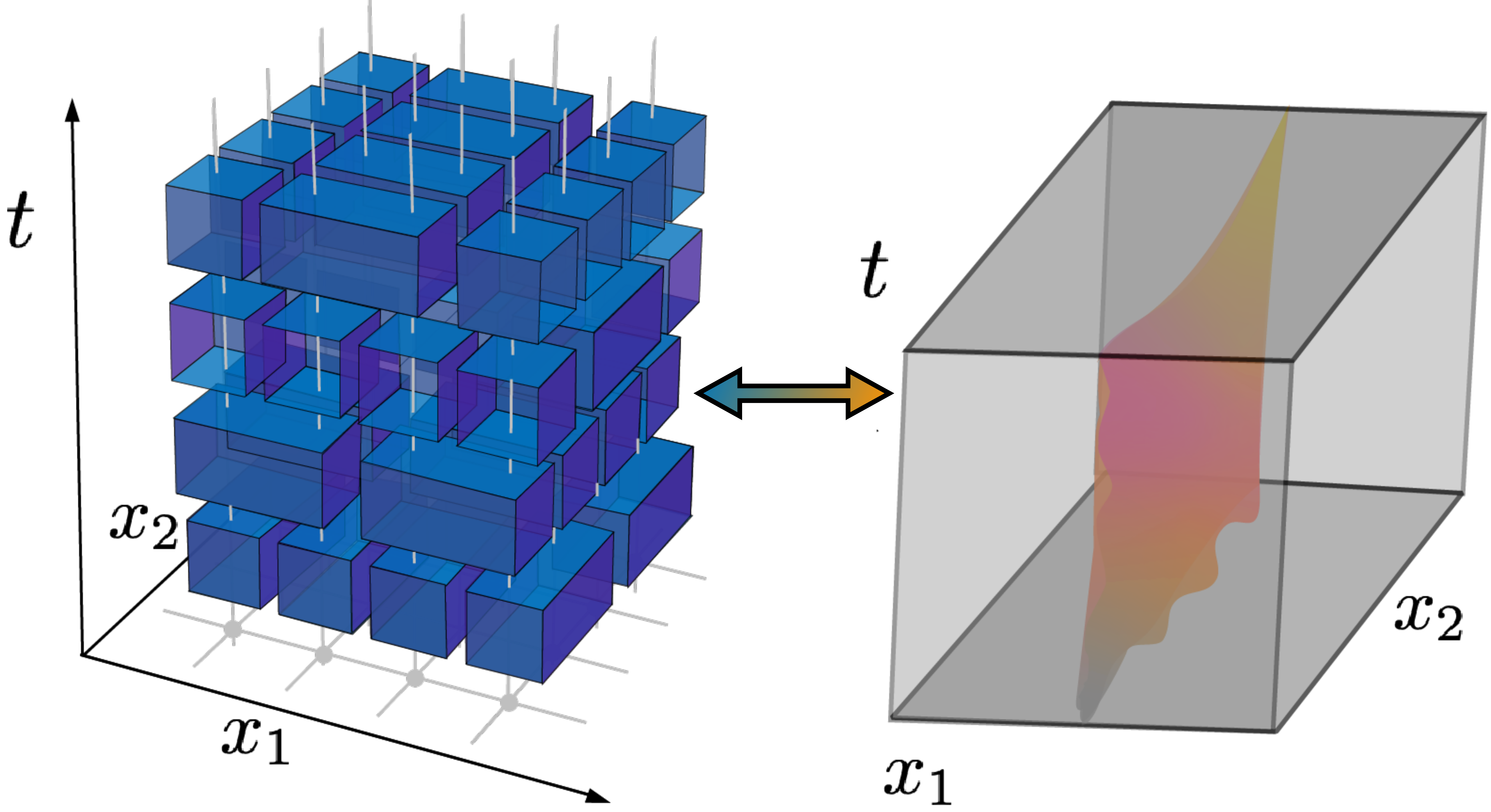}
    \begin{tabular}{c|cc|cc|c}
        \hline \hline
         & \multicolumn{2}{c}{Entanglement} &   \multicolumn{2}{c}{Membrane} & \multicolumn{1}{c}{RG} \\
        \cline{2-3} \cline{4-6}
        {$d+1$} & {$\theta$} & {$\zeta$} & {$\theta$} & {$\zeta$} &  {$\zeta$}\\
        \hline
        2+1 & 0.861(6) & 0.435(33) & 0.84(3) & 0.41(1) & 0.423 \\
        3+1 & 1.401(9) & 0.3(1) & 1.45(4) & 0.22(1)  & 0.211 \\ 
        4+1 & 2.000(5) & - & 2 & -  & -  \\ 
        \hline \hline
    \end{tabular}
    \caption{Cartoon: the entanglement growth of $(d+1)$ dimensional random unitary circuits is conjectured to behave as a $d$-dimensional disordered membrane in a $(d+1)$ elastic manifold. Table: summary of the manuscript results (Entanglement) and comparison with the best numerical estimates~\cite{middleton1995numericalresultsfor,alava1996disorderinducedrougheningin} for random bond Ising model (Membrane), and with the renormalization group analysis~\cite{fisher1986interfacefluctuationsin,halpinhealy1990disorderinducedrougheningof,chauve2001renormalizationofpinned,ledoussal2004functionalrenormalizationgroup,husemann2018fieldtheoryof} (RG). 
    Our results on entanglement propagation are compatible with predictions for membranes pinned by disorder in $(d+1)$ elastic manifolds.  }
    \label{fig:cartoon}
\end{figure}

This work positively answers this question by investigating $(d+1)$ dimensional random Clifford circuits with $2\leq d \leq 4$. The Gottesman-Knill theorem~\cite{nielsen00,gottesman1996classofquantum,gottesman1998theheisenbergrepresentation,aaronson2004improvedsimulationof}, combined with efficient entanglement entropy calculation~\cite{hamma2005bipartiteentanglementand,hamma2005groundstateentanglement}, allows us to perform large-scale numerical simulations for systems 
with up to $N=131072$ qubits. 
We demonstrate that the universal properties implied by the minimal membrane picture are encoded in the growth exponent of the entanglement fluctuations, as summarized in Table~\ref{fig:cartoon}. We also provide robust numerical evidence that the upper critical dimension is ${d=d_c=4}$, beyond which the fluctuations become Gaussian. 

\begin{figure*}[t!]
    \centering
    \includegraphics[width=1\linewidth]{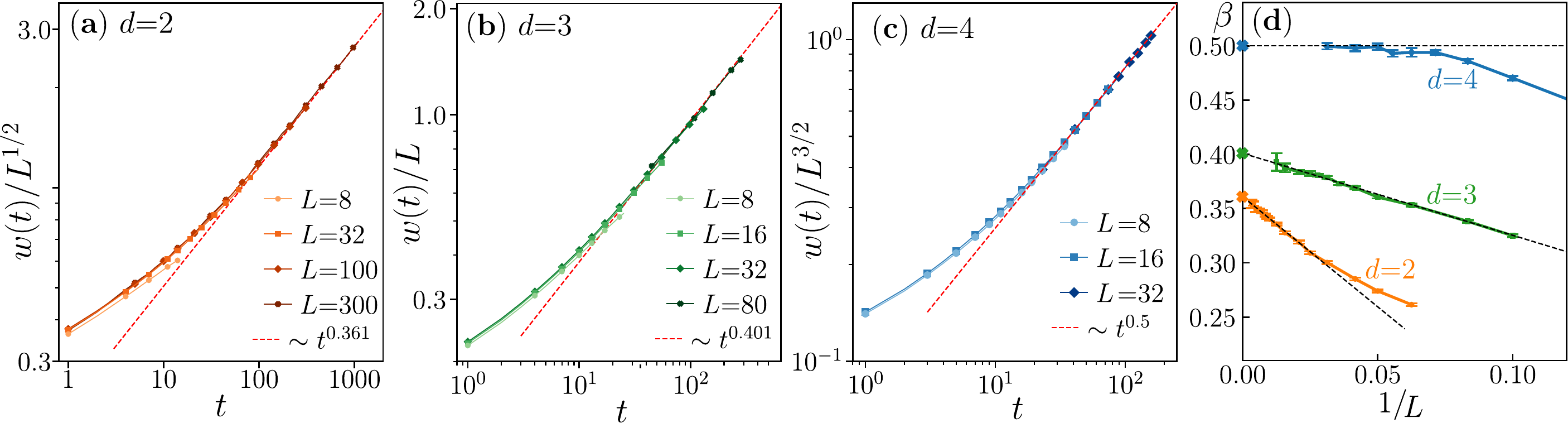}
    \caption{Dynamics of entanglement entropy fluctuations in ($d$+1) dimensional random Clifford circuits. The standard deviation of entanglement entropy $w(t)$ rescaled by $L^{(d-1)/2}$ for system size $L$ is shown, as a function of time $t$, in panel (\textbf{a}),  (\textbf{b}), (\textbf{c}) respectively for $d=2,3,4$. The dashed lines denote a power-law increase $t^{\beta}$ where $\beta$ is an extrapolated exponent. To perform the extrapolation, we fit $w(t)$ for a given $L$ with an algebraic dependence $t^{\beta_L}$ in time interval $[t_1,t_2]$,  chosen in such a way that the value of $\beta_L$ is maximized and $t_2-t_1>40$. The values of $\beta_L$ are shown as functions of $1/L$ in  (\textbf{d}) for $d=2,3,4$, the extrapolation to the $L\to\infty$ limit is performed with a first order polymomial in $1/L$, yielding $\beta=0.361(6)$, $\beta=0.401(9)$,  $\beta=0.500(5)$ respectively for $d=2,3,4$.
    }
    \label{fig2}
\end{figure*}
\textit{Entanglement growth and the membrane conjecture.---} 
We are interested in the dynamics of a random unitary circuit starting from a weakly entangled state $|\Psi_0\rangle$ on the lattice $\Lambda$. Each layer $\hat{\mathcal{U}}=\left(\prod_{\langle \vec{i},\vec{j}\rangle \in I^{(2)}_t} \hat{U}_{\vec{i},\vec{j}}\right) \prod_{\vec{i}\in I^{(1)}_t} \hat{U}_{\vec{i}}$ is composed out of random two-body gates $\hat{U}_{\vec{i},\vec{j}}$ acting on neighboring sites $\vec{i}, \vec{j}$ and on-site random gates $\hat{U}_{\vec{i}}$ applied to random sites belonging to the sets of indices $I^{(1,2)}_t$. 
We quantify the entanglement via the von Neumann entropy, defined for a bipartition $A\cup B$ of the lattice $\Lambda$ and the state $|\Psi_t\rangle = \hat{\mathcal{U}}^t|\Psi_0\rangle$ at time (circuit depth) $t$ as $S(A) \equiv -\mathrm{tr}(\rho_A\log_2\rho_A)$ with $\rho_A = \mathrm{tr}_B(|\Psi_t\rangle\langle\Psi_t|)$. 
The minimal membrane picture conjectured in Ref.~\cite{nahum2017quantumentanglementgrowth} states that, at a coarse-graining level and for the region $A$ bounded by the $(d-1)$ dimensional surface $\partial A$, the entanglement entropy at time $\tau$ is given by 
\begin{equation}
    S_\mathrm{mc}(A,\tau) = \min_\Sigma  \left(S(A',0) + \int_\Sigma dt d^{d-1}x \mathcal{E}(\vec{v})\right).\label{eq:minmembrane}
\end{equation}
The minimization is over the $d$-dimensional membrane $\Sigma(x,t)$ with boundary conditions $\Sigma(x,\tau) = \partial A$ and $\Sigma(x,0)= \partial A'$, and involves the region $A'$ describing the entanglement of the initial state and a \emph{local} velocity $\vec{v}$ defined using the tangent vectors with the minimal angle with respect to the time axis. 
We note that the dependence of the energy density on the local velocity in $\mathcal{E}(\vec{v})$ is a strong assumption in light of the nonlocal nature of $S(A)$.

Without fluctuations, by dimensional analysis, the minimal membrane hypersurface in Eq.~\eqref{eq:minmembrane} scales as $|\partial A| \tau$, compatibly with the leading order average entanglement in Ref.~\cite{znidari2020entanglementgrowthin}.
The randomness of the circuit is reflected in the disordered nature of $\Sigma(x,t)$. Assuming uncorrelated and Gaussian noise, $\Sigma(x,t)$ is fixed as the ground state of a classical Hamiltonian~\cite{nattermann1985isingdomainwall,huse1985pinningandroughening,fisher1986interfacefluctuationsin} 
\begin{equation}
    \mathcal{H}[\Sigma(x,t)] = \int dt d^{d-1}x \left( \frac{\kappa}{2} (\nabla \Sigma(x,t))^2 + V(x)\right),
\end{equation}
with $V(x)$ a Gaussian noise with zero mean $\overline{V(x)} = 0$ and variance $\overline{V(x)V(x')} = \gamma \delta(x-x')$, for some parameters $\kappa$ and $\gamma$. This problem is fully characterized by the exponents $\theta$ and $\zeta$, related by the hyperscaling relation $2\zeta = \theta + d-2$~\cite{middleton1995numericalresultsfor,wiese2022theoryandexperiments} and governing the energy and width fluctuations, respectively
\begin{equation}
    \begin{split}
        E(x,t) &\equiv \sqrt{\langle \mathcal{H}[\Sigma(x,t)]^2\rangle-\langle \mathcal{H}[\Sigma(x,t)]\rangle^2} \propto L^\theta, \\
        W(x,t) &\equiv \sqrt{\langle \Sigma(x,t)^2\rangle-\langle \Sigma(x,t)\rangle^2} \propto L^{\zeta}.
    \end{split}
\end{equation}
Renormalization group and numerical simulations predict $\zeta\simeq 0.21(4-d)$ and $\theta=0.84(3)$ ($\theta=1.45(4)$) in $d=2$ ($d=3$), while the fluctuations become Gaussian at the upper critical dimension $d_c=4$~\cite{fisher1986interfacefluctuationsin}.

The minimal membrane picture conjecture, cf. Eq.~\eqref{eq:minmembrane}, provides a link between the fluctuations of the membrane $\Sigma(x,t)$ and of the entanglement entropy. 
In particular, 
entanglement entropy for a fixed realization is given by~\cite{nahum2017quantumentanglementgrowth}
\begin{equation}
    S(t) = \tilde{v} t |\partial A| t + \tilde{b} t^{\theta + 1 - d} \chi(t) 
    ,
    \label{eq:ent}
\end{equation}
where $\chi(t)$ is a stochastic variable, $\tilde{v}$ and $\tilde{b}$ are two constants depending on the microscopic details of the model, and we neglected sub-subleading contributions.
For concreteness, from now on, we consider $\Lambda$ to be a hyperrectanglular $L_1\times\dots \times L_d$ lattice, with $L_1 = L$ and $L_{2}=\ldots=L_{d}=L/2$, assuming periodic boundary conditions in all directions.
As the subsystem $A$ we take a hyperrectangular sublattice of $\Lambda$ of size $x \times L_2 \times \ldots \times L_d$.
Then, the membrane growth exponents imply that the spatiotemporal fluctuations obey
\begin{equation}
\begin{split}
     w(x,t) &\equiv \sqrt{\langle S(x,t)^2\rangle-\langle S(x,t)\rangle^2} \propto L^{(d-1)/{2}}t^{\beta},\\
        G(r,t) &\equiv \sqrt{\langle (S(x,t) - S(x+r,t))^2\rangle} =  r^{\beta/\zeta} g\left(\frac{r}{t^\zeta}\right),
\end{split}\label{eq:wxt}
\end{equation}
where $\beta=\theta + (1-d)/2$ is the growth exponent, $\langle\circ\rangle$ is the disorder average and $g$ is a scaling function.
In the following, we numerically extract the growth exponent $\beta$ (and hence $\theta$) and $\zeta$ by studying the entanglement entropy fluctuations in random circuits and comparing them with the results for disordered elastic manifolds~\cite{wiese2022theoryandexperiments}. 

\begin{figure*}[t!]
    \centering
    \includegraphics[width=1\linewidth]{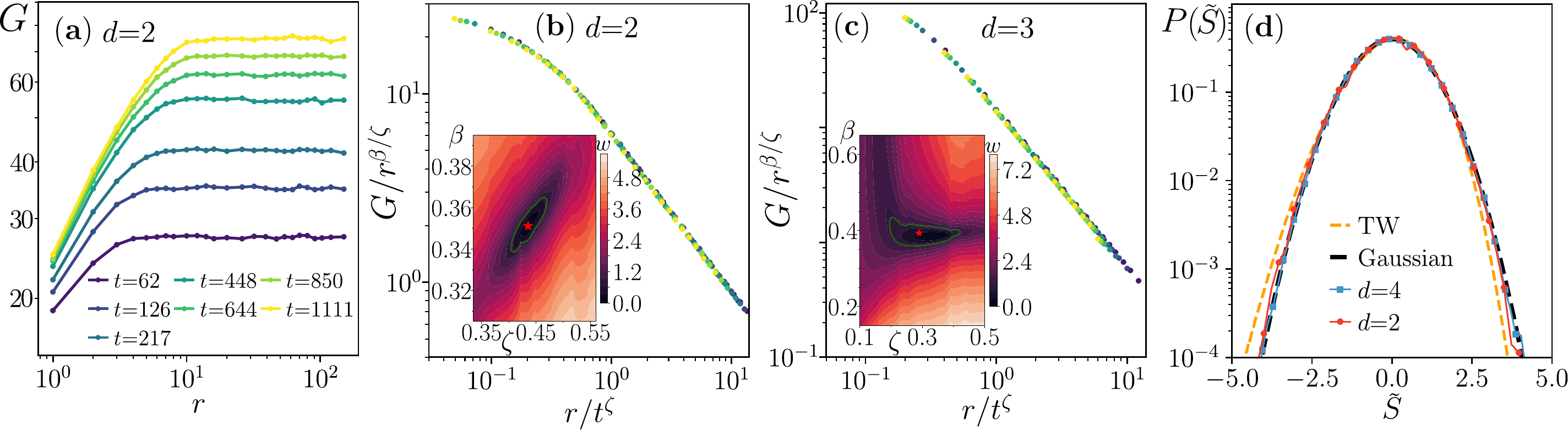}
    \caption{ Spatiotemporal fluctuations of entanglement entropy in random Clifford circuits. The entanglement entropy correlation $G(r,t)$ is plotted in (\textbf{a}) as a function of $r$ for various times $t$ for the (2+1) dimensional circuit of size $L=360$. The data for different times $t$ are collapsed according to \eqref{eq:wxt} in panel (\textbf{b}), the optimal obtained values of the exponents are $\beta=0.351(12)$, $\zeta = 0.435(33)$, the cost function $w$ is shown in the inset. The same is shown for $d=3$ in panel (\textbf{c}), the optimal exponents are $\beta=0.39(3)$, $\zeta = 0.3(1)$. Distribution of $\tilde S$ -- the rescaled entanglement entropy, at time $t=L$ is shown in (\textbf{d}) for $(2+1)$ and $(4+1)$ dimensional circuits and compared with Gaussian and standardized Tracy-Widom distributions (dashed lines). 
    }
    \label{fig3}
\end{figure*}

\textit{Numerical results.---}
Assuming the membrane picture conjecture, the above discussion is relevant for any choice of the unitary ensemble of the random gates. Nevertheless, numerical simulations of generic (Haar) random circuits requires computational resources 
that scale exponentially with the number of investigated qubits.
Therefore, this work resorts to Clifford circuits whose dynamics for stabilizer states on $N$ qubits are efficiently simulable via the Gottesmann-Knill theorem~\cite{gottesman1996classofquantum,nielsen00,gottesman1998theheisenbergrepresentation,aaronson2004improvedsimulationof}.

Denoting the Pauli strings as $\hat{P} = e^{i\phi}\prod_{\vec{i}\in \Lambda} \hat{X}^{n_{\vec{i}}}_{\vec{i}}\hat{Z}^{m_{\vec{i}}}_{\vec{i}}$, any pure stabilizer state $|\Psi\rangle$ is determined by $N= |\Lambda|$ independent Pauli strings $\hat{P}_\mu$ as $|\Psi\rangle\langle\Psi| =\prod_{\mu=1}^N ({\hat\openone+\hat{P}_\mu})/{2}$.
Hence, the stabilizer state is encoded in a $N\times (2N+1)$ matrix $M  =(\phi^\mu|n^\mu_{\vec{i}}|m^\mu_{\vec{i}})$ with elements in $0,1$ and whose rows define the stabilizing Pauli strings. Clifford gates are defined as unitary operators that map a Pauli string into a \emph{single} Pauli string.  
Hence, the action of Clifford gates can expressed as operations in the $\mathbb{Z}_2$ field on the matrix $M$. It follows that the unitary evolution under Clifford circuits is simulable in resources scaling polynomially with the number of qubits $N$. 
Furthermore, the entanglement entropy of stabilizer states is efficiently computable as $S(A) = r_A - N_A$, where $r_A = \mathit{rk}_2 M_A$ is the rank of the submatrix $M_A = (n_{\vec{i}}^\mu|m_{\vec{i}}^\mu)_{\vec{i}\in A}$ and $N_A$ is the number of qubits in $A$~\cite{nahum2017quantumentanglementgrowth}.
We note that earlier simulations in the literature were based on standard Jordan-Gauss elimination, with complexity $O(N^3)$. Here, 
to calculate entanglement entropy we employ an 
asymptotically fast rank-revealing algorithm with complexity $O(N^3 / \log_2 N)$~\cite{andren2007onthecomplexity,albrecht2011efficient}.  
We combine an efficient implementation of this algorithm~\cite{Bertolazzi14} with a state-of-the-art \textsc{STIM}~\cite{stim} library to perform large-scale computations for systems of up to $N=131072$ qubits. This allows us to calculate entanglement entropy for the largest considered $N$ around two-orders of magnitude faster than a naive implementation of Jordan-Gauss elimination in \textsc{STIM}.

In our implementation, $I^{(1)}_t$ is a set of $N/2$ sites at which the single qubit unitaries act. The one body gates are selected randomly, with equal probability, to be either Hadamard, $H$, or phase, $\sqrt{Z}$, gates. 
Instead, $I^{(2)}_t$ is a set of $N/4$ pairs of nearest-neighboring sites $\langle \vec{i},\vec{j}\rangle$ where $\vec{i}$ is a site picked randomly with uniform distribution on $\Lambda$ and $\vec{j}=\vec{i}+\vec{e}_u$ where $\vec{e}_u$ is a versor in a randomly chosen direction $u=1,\dots,d$. 
The two-body gates acting on $I^{(2)}_t$ are chosen, with equal probability, between CNOT gates controlled on the left or right argument~\cite{nielsen00}.
For concreteness, we consider the fully polarized initial state $|\Psi_0\rangle = \otimes_{\vec{i}\in \Lambda} |0\rangle$. We have tested other low-entangled initial states finding quantitatively consistent results.

First, we consider the fluctuations $w(t)\equiv w(L/2,t)$ of the entanglement entropy, cf. Eq.~\eqref{eq:wxt}. Our findings are summarized in Fig.~\ref{fig2}, where we let the system evolve up to time $t=4L$. To extrapolate the growth exponent $\beta_L$ for a fixed system size $L$, we fit the numerical results with an algebraic dependence $t^{\beta_L}$ in the time interval $[t_1,t_2]$. We consider $t_2-t_1>40$, chosen in order to maximize $\beta_L$. We extrapolate the thermodynamic limit $L\to\infty$ using a first-order polynomial in $1/L$, obtaining respectively $\beta = 0.361(6)$, $\beta=0.401(9)$ and $\beta=0.500(5)$ for $d=2,3,4$ in agreement with the exponents obtained for the pinned membrane, c.f. Table~\ref{fig:cartoon}, where the hyperscaling relation is used to translate the obtained values of $\beta$ to the exponents $\theta$.

Furthermore, we study the spatiotemporal fluctuations $G(r,t)$, cf. Eq.~\eqref{eq:wxt}. We fix the system sizes to be equal to $L=360$ and $L=64$ for $d=2$ and $d=3$, respectively, and study the growth of the fluctuations varying $r$ and $t$. (The $d=2$ case is presented in Fig.~\ref{fig3}(a), while the result for $d=3$ is qualitatively similar and are not shown here.)
The spatiotemporal dependence of $G(r,t)$ can be analyzed with the hypothesis~\eqref{eq:wxt}, where the growth exponents $\beta$ and $\zeta$ obtained by rescaling of results for different circuit depths $t$ with the aim of finding an optimal collapse the data onto a single curve.
To perform the analysis in an unbiased fashion, we consider both $\beta$ and $\zeta$ as unknown parameters, and optimize the cost function $C(\beta,\zeta)= \sum_{i=2}^{n-1} c(x_i,y_i,d_i)$ with  $x_i=(r_i/t_i^\zeta)$, $y_i=G(r_i,t_i)/r_i^{\beta/\zeta}$ and $d_i=\sigma(y_i)$ respectively the hyperparameters, data, and standard deviations. The indices $i=0,\dots,\mathcal{N}_\mathrm{data}$ are ordered such that $x_i<x_j$ when $i<j$~\cite{zabalo2020criticalpropertiesof,sierant2022measurementinducedphase}, and the density $c(x_i,y_i,d_i) = (y-\overline{y})^2/\Delta$ depends on $\overline{y} = \left[(x_{i+1}-x_i)y_{i-1} - (x_{i-1}-x_i)y_{i+1}\right] /(x_{i+1}-x_{i-1})$ and on 
\begin{equation}
    \Delta = d_i^2 + \left(\frac{x_{i+1}-x_{i}}{x_{i+1}-x_{i-1}}d_{i-1}\right)^2 + \left(\frac{x_{i-1}-x_{i}}{x_{i+1}-x_{i-1}}d_{i+1}\right)^2.\nonumber 
\end{equation}
The optimal growth exponents are obtained via minimization $\beta^\star,\zeta^\star = \arg\min C(\beta,\zeta)$, and lead to the data collapses in Fig.~\ref{fig3}(b) and (c) for $d=2$ and $d=3$ respectively. In the insets, we present the cost function landscape $C(\beta,\zeta)$. In $d=2$, our estimates $\beta=0.351(12)$ and $\zeta=0.435(33)$ correspond to a relatively narrow dip in the cost function leading to exponents that are compatible with the pinned membrane ones. 
In $d=3$, the cost function, cf.~Fig.~\ref{fig3}(c,inset), possesses a wide minimum which leads to larger error bars on the estimates $\beta=0.39(3)$ and $\zeta=0.3(1)$. Still, these values are compatible with the membrane predictions, cf. Table~\ref{fig:cartoon}.

Lastly, we study the distribution of the variable $\tilde{S} = (S-\langle{S}\rangle)/\sigma(S) $, with $\sigma(S)$ the standard deviation and $\langle S\rangle$ the average. In one dimension, this variable follows the centralized Tracy-Widom (TW) distribution~\cite{nahum2017quantumentanglementgrowth}, whereas, for $d=d_c=4$, the system is expected to reach a Gaussian distribution~\cite{fisher1986interfacefluctuationsin,wiese2022theoryandexperiments}. 
For $d=2,3$ the distribution is unknown but is expected to differ from the TW and Gaussian ones. Our results for $d=2,4$ are given in Fig.~\ref{fig3}(d) at $t=L$ for $L=320$ and $L=32$, respectively. We see that both distribution are closer to the Gaussian fit than to the TW. To quantify the closeness to Gaussianity, we investigate the skewness $\kappa_3$ of $P(\tilde{S})$. The TW distribution relevant for the $d=1$ case yields $\kappa_3\approx0.2241$, while we find that the $d=2$ distribution has $\kappa_3 = 0.07(1)$, showing that the distribution is nongaussian. For $d=3$, we obtain that $\kappa_3 = 0.004(3)$. Hence, while the distribution $P(\tilde{S})$ approaches the Gaussian distribution with increasing system dimension, we still observe mild deviations from gaussianity at the level of the $P(\tilde{S})$ for $d=3$, consistently with the non-mean field values of the exponents $\theta$ and $\zeta$. Instead, for $d=4$, the numerically obtained $P(\tilde{S})$ that is indistinguishable from Gaussian, and we find $\kappa_3 = 0.000(4)$ confirming the membrane picture prediction for the upper critical dimension. 

\textit{Discussion and conclusion.---} 
This manuscript investigates the structural properties of entanglement propagation in  $(d+1)$ dimensional random unitary circuits, as encoded in the fluctuations and distribution of the entanglement entropy. 
Focusing on Clifford circuit implementations, we compute the growth exponents $\theta$ and $\zeta$ for $d\le 4$ with high accuracy. We find they are compatible with the corresponding exponents governing the fluctuations of a membrane pinned by disorder in an elastic manifold, cf. Table~\ref{fig:cartoon}. 
Furthermore, at $d=4$, the entanglement fluctuations become Gaussian, in accordance with the upper critical dimension of membranes pinned by disorder in $d+1$ manifolds. 
Overall, our findings demonstrate the effectiveness of the recently proposed minimal cut (membrane) conjecture in capturing the essential structure of entanglement dynamics~\cite{nahum2017quantumentanglementgrowth}. 

Beyond numerical considerations, suitable limits (e.g., large local Hilbert space dimension) may be analytically treatable for Haar and Clifford circuits even in higher dimensions~\cite{zhou2019emergentstatisticalmechanics, zhou2020entanglementmembranein,li2021statisticalmechanicsmodel}. 
At the same time, it would be interesting to investigate the properties of the volume-law phase in $(d+1)$ monitored systems~\cite{li2018quantumzenoeffect,li2019measurementdrivenentanglement,skinner2019measurementinducedphase,zabalo2022operatorscalingdimensions,zabalo2020criticalpropertiesof,sierant2022universalbehaviorbeyond,weinstein2022measurementinducedpower,turkeshi2020measurementinducedcriticality,turkeshi2022measurementinducedcriticality,sierant2022measurementinducedphase,weinstein2022scramblngtransitionin,lunt2022quantumsimulationusing,kelly2022coherencerequirementsfor,szyniszewski2019entanglementtransitionfrom,szyniszewski2020universalityofentanglement,jian2021measurementinducedphase,lopezpiqueres2020meanfieldentanglement,vasseur2019entanglementtransitionsfrom,jian2020measurementinducedcriticality,nahum2921measurementandentanglement,nahum2023renormalizationgroupfor,bao2020theoryofthe,choi2020quantumerrorcorrection} and long-range systems~\cite{sierant2022dissipativefloquet,richter2023transportandentanglement,sharma2022measurementinducedcriticality,block2022measurementinducedtransition,sierant2022controllingentanglementat}. While recent works for $(1+1)D$ circuits  revealed a rich structure related to polymer defects~\cite{li2021statisticalmechanicsof,vijay2020measurementdrivenphase,fan2021selforganizederror,li2021entanglementdomainwalls}, the case of higher dimensions has been left essentially unexplored. A first step in $(d+1)$ dimension was recently presented by Ref.~\cite{zhu2023structured} for interacting Majorana circuits, where nontrivial subleading corrections due to the topological nature of the system were found. An interesting idea is to relate the $L\ln L$ subleading behavior to Fermi, surface corrections in the spirit of the Widom conjecture~\cite{swingle}, or to the statistical mechanics of loop models~\cite{nahum2011,nahum2015,nahum2019,klocke2023majorana}.
We leave these questions for future investigation.

\begin{acknowledgments}
\textit{Acknowledgments.---}
We thank A. Nahum and T. Zhou for discussions. 
We acknowledge the workshop "Dynamical Foundation of Many-Body Quantum Chaos" at Institute Pascal (Orsay, France) for hosting P.S. and X.T. during the finalizing stage of this manuscript's writing.
X.T. and M. S. acknowledge support from the ANR grant “NonEQuMat”
(ANR-19-CE47-0001) and computational resources on the Coll\'ge de France IPH cluster. 
P.S. and M. L. acknowledge support from: ERC AdG NOQIA; Ministerio de Ciencia y Innovation Agencia Estatal de Investigaciones (PGC2018-097027-B-I00/10.13039/501100011033, CEX2019-000910-S/10.13039/501100011033, Plan National FIDEUA PID2019-106901GB-I00, FPI, QUANTERA MAQS PCI2019-111828-2, QUANTERA DYNAMITE PCI2022-132919, Proyectos de I+D+I “Retos Colaboración” QUSPIN RTC2019-007196-7); MICIIN with funding from European Union NextGenerationEU(PRTR-C17.I1) and by Generalitat de Catalunya; Fundació Cellex; Fundació Mir-Puig; Generalitat de Catalunya (European Social Fund FEDER and CERCA program, AGAUR Grant No. 2021 SGR 01452, QuantumCAT \ U16-011424, co-funded by ERDF Operational Program of Catalonia 2014-2020); Barcelona Supercomputing Center MareNostrum (FI-2023-1-0013); EU (PASQuanS2.1, 101113690); EU Horizon 2020 FET-OPEN OPTOlogic (Grant No 899794); EU Horizon Europe Program (Grant Agreement 101080086 — NeQST), National Science Centre, Poland (Symfonia Grant No. 2016/20/W/ST4/00314); ICFO Internal “QuantumGaudi” project; European Union’s Horizon 2020 research and innovation program under the Marie-Skłodowska-Curie grant agreement No 101029393 (STREDCH) and No 847648 (“La Caixa” Junior Leaders fellowships ID100010434: LCF/BQ/PI19/11690013, LCF/BQ/PI20/11760031, LCF/BQ/PR20/11770012, LCF/BQ/PR21/11840013). Views and opinions expressed are, however, those of the author(s) only and do not necessarily reflect those of the European Union, European Commission, European Climate, Infrastructure and Environment Executive Agency (CINEA), nor any other granting authority. Neither the European Union nor any granting authority can be held responsible for them. 
\end{acknowledgments}


%

\end{document}